\documentclass[twoside]{article}
\usepackage{fleqn,espcrc2}
\usepackage{amsmath}

\usepackage{epsf}

\newcommand{\be}{\begin{eqnarray}}
\newcommand{\ee}{\end{eqnarray}}
\newcommand{\bdm}{\begin{displaymath}}
\newcommand{\edm}{\end{displaymath}}
\newcommand{\<}{\langle}
\renewcommand{\>}{\rangle}

\def\spose#1{\hbox to 0pt{#1\hss}}
\def\ltapprox{\mathrel{\spose{\lower 3pt\hbox{$\mathchar"218$}}
 \raise 2.0pt\hbox{$\mathchar"13C$}}}
\def\gtapprox{\mathrel{\spose{\lower 3pt\hbox{$\mathchar"218$}}
 \raise 2.0pt\hbox{$\mathchar"13E$}}}
\def\inapprox{\mathrel{\spose{\lower 3pt\hbox{$\mathchar"218$}}
 \raise 2.0pt\hbox{$\mathchar"232$}}}

\title{Exact Chiral Symmetry for Domain Wall Fermions with Finite
$L_s$\thanks{Presented by U.~M.~Heller at {\sl Lattice 2000}}
}

\author{Robert G. Edwards\address{Jefferson Lab, 12000 Jefferson Avenue,
                                  MS 12H2, Newport News, VA 23606, USA}
        and
        Urs M. Heller\address{CSIT, Florida State University,
                              Tallahassee, FL 32306-4120, USA}
}

\begin{document}

\begin{abstract}
We show how the standard domain wall action can be simply modified to
allow arbitrarily exact chiral symmetry at finite fifth dimensional
extent $L_s$. We note that the method can be used for both quenched and
dynamical calculations.  We test the method using smooth and
thermalized gauge field configurations.  We also make comparisons of
the performance (cost) of the domain wall operator for spectroscopy
compared to other methods such as the overlap-Dirac operator and find
both methods are comparable in cost.
\end{abstract}

\maketitle

\section{The problem with small eigenvalues}

Recently a great deal of theoretical progress has been made in the
construction of lattice regularizations of fermions with good chiral
properties \cite{Kaplan,overlap}. For use in practical numerical
simulations, though, approximations to these formulations are necessary.
In the formulation using domain wall fermions (DWF)~\cite{Kaplan,Shamir},
the extent of an auxiliary fifth dimension has to be kept finite in
numerical simulations while chiral symmetry holds strictly only in the
limit of infinite fifth dimension. The violations of chiral symmetry are
expected to be suppressed exponentially in the extent of the fifth
dimension~\cite{Shamir,finite_Ls}, but in practice the coefficient in the
exponent can be quite small~\cite{Columbia,dwf_problems} and the
suppression correspondingly slow.

In the case of overlap fermions there is no such problem in principle.
However, there is a problem of practicality: how to deal efficiently
with $\epsilon(H) = H / \sqrt{H^2}$, where $H$ is some auxiliary Hermitian
lattice Dirac operator for large negative mass, but free of doublers.
Most commonly, the Hermitian Wilson-Dirac operator $H_w$ is used.

Any good numerical approximation to $\epsilon(H)$ must retain the
property $\epsilon(H)^2=1$ which is crucial for the Ginsparg-Wilson
relation
\be
\{ \gamma_5, D_{ov}(0) \} = 2 D_{ov}(0) \gamma_5 D_{ov}(0)
\ee
to hold and exact chiral symmetry for the overlap Dirac
operator~\cite{Herbert}
\be
D_{ov}(m) = \frac{1}{2} \left[ 1 + m + (1-m) \gamma_5 \epsilon(H) \right]
\ee
at $m=0$ to be preserved. A good approximation of $\epsilon(x)$ is
difficult to achieve for small $x$.

\vskip -12cm
\rightline{FSU-CSIT-00-27}
\rightline{JLAB-THY-00-34}
\vskip +11.2cm

Unfortunately, at least in quenched simulations with the Wilson gauge action,
$H_w(M) = \gamma_5 D_w(-M)$, with $D_w(M)$ the Wilson-Dirac operator,
has a non-vanishing density of zero eigenvalues, $\rho(0;M)$, for any
gauge coupling $\beta$~\cite{EHN_rho0}. $\rho(0;M)$ decreases quite
rapidly with increasing $\beta$, roughly as $\rho(0;M)/\sigma^{3/2} \ \sim \
e^{-e^\beta}$~\cite{EHN_rho0}, but at currently used couplings a
considerable number of modes with small eigenvalues exists.

In all numerical methods to implement overlap fermions one can enforce
accuracy of the approximation to $\epsilon(H_w)$ by projecting out the
lowest few $H_w$ eigenvectors and adding their correct contribution
exactly (see Ref.~\cite{ParalComp}): with $H_w v_i = \lambda_i v_i$,
\begin{eqnarray}
\epsilon(H_w) &\!\!\!\!\!=&\!\!\!\!\! \sum_{i=1}^n |v_i\>
 \epsilon(\lambda_i) \<v_i| + {\cal P}_\perp^{(n)}
 {\rm App}[\epsilon(H_w)] {\cal P}_\perp^{(n)} , \nonumber \\
{\cal P}_\perp^{(n)} &\!\!\!\!\!=&\!\!\!\!\! {\bf 1} -
 \sum_{i=1}^n |v_i\> \<v_i| .
\end{eqnarray}
The number $n$ of projected eigenvectors is chosen such that the
approximation ${\rm App}[\epsilon(H_w)]$ to $\epsilon(H_w)$ is
sufficiently accurate in the subspace spanned by ${\cal P}_\perp^{(n)}$.

\section{Implications for domain wall fermions}

Integrating out the heavy fermions, one can show that the effective Dirac
operator for the physical 4-d fermions from the domain wall fermion
approach is~\cite{Herbert_DWF,EH_5d}
\be
D_{tov}(m) = \frac{1}{2} \Bigl[1+m + (1-m) \gamma_5
\varepsilon_{L_s/2}(H_T) \Bigr]
\ee
with $\varepsilon_N(x)$ the polar decomposition approximation to
$\epsilon(x)$~\cite{polar}
\be
\varepsilon_N(x) &\!\!\!\!\!=&\!\!\!\!\! \frac{(1+x)^{2N} -
 (1-x)^{2N}}{(1+x)^{2N} + (1-x)^{2N}} \nonumber \\
&\!\!\!\!\!\approx&\!\!\!\!\! \tanh(2Nx)
\label{eq:polar}
\ee
and~\cite{borici}
\be
H_T(M) = H_w(M) \frac{1}{2 + \gamma_5 H_w(M)} .
\label{eq:H_T}
\ee
Zero modes of $H_w(M)$ are also zero modes of $H_T(M)$!

Small eigenvalues of $H_T$ induce chiral symmetry violations, unless
$L_s$ is very large --- $\tanh(y)$ is within about $10^{-5}$ of 1
for $y \ge 6.1$. And the density of small eigenvalues of $H_T$ is the same
as the density of small eigenvalues of $H_w$!

\begin{figure}
\epsfxsize=2.5in
\centerline{\epsfbox[50 50 600 570]{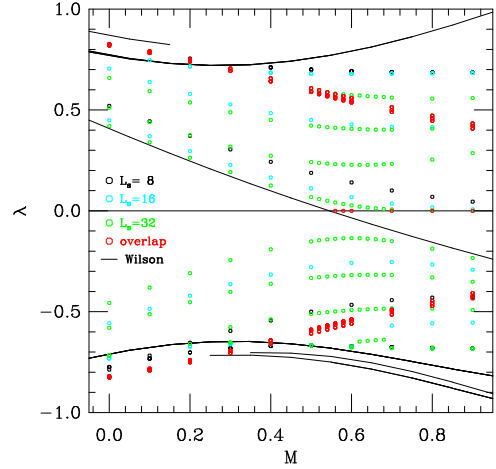}}
\vskip -4mm
\caption{Eigenvalues of the hermitian domain wall, Wilson and overlap
operators as function of the negative Wilson mass value $M$ on a single
instanton configuration.}
\vskip -4mm
\label{fig:flow1}
\end{figure}

Manifestations of the induced chiral symmetry breaking are, for example,
violations of the Gellmann-Oakes-Renner relation
\be
m \chi_\pi \equiv m \< \sum_x \pi(x) \pi(0) \> = \< \bar \psi \psi \> ~,
\label{eq:GMOR}
\ee
finding a non-zero pion mass for massless quarks, and absence of exact
zero modes in topologically non-trivial gauge field backgrounds.
This latter is illustrated in Fig.~\ref{fig:flow1} where we show the
spectral flow of the hermitian domain wall operator $H_{DW}(0;M) = \gamma_5
{\cal J} D^{(5)}_{DW}(0;M)$, with ${\cal J}$ the inversion operator
of the fifth direction, the hermitian Wilson operator $H_w(M)$ and the
hermitian overlap $H_{ov}(0;M) = \gamma_5 D_{ov}(0:M)$ on a single instanton
configuration. Right after a Wilson eigenvalue crosses zero the overlap
operator has an exact zero mode while the lowest eigenvalue of the domain
wall operator is clearly non-zero even for $L_s=32$ for $M$ considerably
after the crossing.

\section{Exact chiral symmetry at finite $L_s$}

With finite lattice spacing in the $5^{th}$ direction, $a_5$, the domain
wall fermion (DWF) action is
\be
S_{DW} = - \bar \Psi D^{(5)}_{DW} \Psi ~.
\ee
The kernel of the 5-d operator is, including the boundary conditions
used to introduce an explicit quark mass $m$,
\be
&\!\!\!\!\!\!&\!\!\!\!\!\! D^{(5)}_{DW} = \\
&\!\!\!\!\!\!&\!\!\!\!\!\! \begin{pmatrix}
D_+ - \hat{A}P_- & -P_- & \cdots & 0 & (m - \hat{A}) P_+ \cr
-P_+ & D_+ & \cdots & 0 & 0 \cr
\vdots & \vdots & \vdots & \vdots & \vdots \cr
0 & 0 & \cdots & D_+ & -P_- \cr
m P_- & 0 & \cdots & -P_+ & D_+ \cr
\end{pmatrix} \nonumber
\ee
where $P_{\pm} = (1 \pm \gamma_5)/2$ and $D_+ = a_5 D_w(-M) + 1$.
$\hat{A}(m)$ is the new ingredient, acting on the light fermions, chosen
to insure chiral symmetry. We integrate out the $L_s-1$ extra fermion
fields and the pseudo-fermions to obtain~\cite{EH_5d}
\begin{eqnarray}
&\!\!\!\!\!\!&\!\!\!\!\!\! D_{tov}(m;H_T) \nonumber \\
&\!\!\!\!\!=&\!\!\!\!\! \left\{ {\cal P}^{-1}
 \left[ D^{(5)}_{DW}(1) \right]^{-1}
 D^{(5)}_{DW}(m) {\cal P} \right\}_{11} \nonumber \\
&\!\!\!\!\!=&\!\!\!\!\! \frac{1}{2} \Bigl[1+m + (1-m) \gamma_5
 \varepsilon_{L_s/2}(a_5 H_T) \Bigr] \\
&\!\!\!\!\!&\!\!\! + \text{term with } \hat{A}(m) ~. \nonumber
\end{eqnarray}
Here $\varepsilon_N(x)$ is the polar decomposition approximation to
$\epsilon(x)$ introduced in (\ref{eq:polar}), and ${\cal P}$ is such
that $({\cal P}^{-1} \Psi)_1 = q$ is the light fermion field,
\begin{eqnarray}
{\cal P}_{jk} = \begin{cases}
 P_- \delta_{j,k} + P_+ \delta_{j+1,k} & \text{for $j < L_s$} \\
 P_- \delta_{L_s,k} + P_+ \delta_{1,k} & \text{for $j = L_s$} ~. \end{cases}
\end{eqnarray}
$H_T$, given in (\ref{eq:H_T}), is the auxiliary Hamiltonian associated with
the transfer matrix $T$ in the $5^{th}$ direction,
\be
T \equiv \frac{1 - a_5 H_T}{1 + a_5 H_T} ~.
\ee

With $H_T v_i= \lambda_i v_i$, $T v_i = T_i v_i$ we set,
\begin{eqnarray}
\hat A(m) = (1-m) \gamma_5 (H_w P_- -1) \sum_{i=1}^n g_i |v_i\> \<v_i| ~, \\
g_i = \frac{1}{2} \left[ - \left( T_i^{-L_s} -1 \right)
 + \left( T_i^{-L_s} +1 \right) \epsilon(\lambda_i) \right] ~.
\end{eqnarray}
This ``projects'' low-lying eigenvalues of $H_T$ from
$\varepsilon_{L_s/2}(a_5 H_T)$ and gives the correct $\epsilon(\lambda_i)$.
Then $D_{tov}(m;H_T)$ becomes $D_{ov}(m;H_T)$,
\be
D_{ov}(m;H_T)=\frac{1}{2} \left[1+m + (1-m) \gamma_5\epsilon(H_T)  \right] ~,
\ee
the overlap operator with auxiliary Hamiltonian $H_T$.

\begin{figure}
\epsfxsize=2.5in
\centerline{\epsfbox[50 50 600 570]{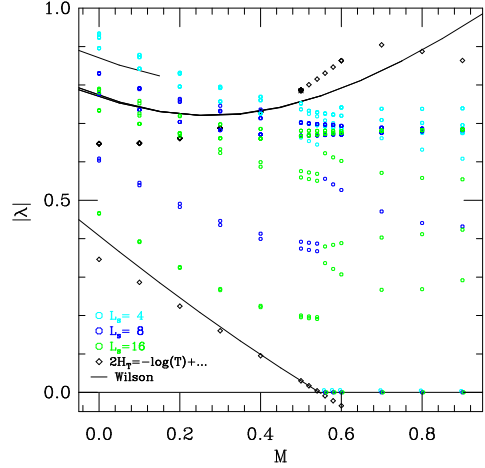}}
\vskip -4mm
\caption{Eigenvalues of the domain wall operator with projection as function
of the negative Wilson mass value $M$ on a single instanton configuration.
Also show is the Wilson operator spectral flow and the eigenvalues of
$2H_T(M)$.}
\vskip -4mm
\label{fig:flow2}
\end{figure}

With projection of the 5 lowest eigenvectors of $H_T$ we now find the
expected zero-modes of the massless projected domain wall operator
$\sqrt{(D^{(5)}_{DW}(0))^\dagger D^{(5)}_{DW}(0)}$ for the same instanton
configuration as in Fig.~\ref{fig:flow1} already for moderate $L_s$, as can
be seen in Fig.~\ref{fig:flow2}.

As already mentioned, the Gellmann-Oakes-Renner relation (\ref{eq:GMOR})
is a sensitive test of chiral symmetry. For overlap fermions
\be
m \< b| (\gamma_5 {\tilde D}^{-1}_{ov}(m))^2 |b \> =
\< b| {\tilde D}^{-1}_{ov}(m) |b \>
\label{eq:GMOR_b}
\ee
holds for any chiral vector $\gamma_5 |b \> = \pm |b \>$, and averaging
over several chiral Gaussian vectors $|b \>$ this becomes a stochastic
estimate for the Gellmann-Oakes-Renner relation (\ref{eq:GMOR}).
In (\ref{eq:GMOR_b}) ${\tilde D}^{-1}_{ov}(m)$ is the external fermion
propagator with the contact term subtracted and multiplicatively
normalized~\cite{ParalComp}, which avoids ${\cal O}(a)$ effects~\cite{OpImpGW},
\begin{eqnarray}
&\!\!\!\!\!\!&\!\!\!\!\!\! (1-m) {\tilde D_{ov}}^{-1}(m)
  = \left[D_{ov}^{-1}(m) - 1 \right] \\
&\!\!\!\!\!=&\!\!\!\!\! \left[\left\{ {\cal P}^{-1}
\left[ D^{(5)}_{DW}(m) \right]^{-1} D^{(5)}_{DW}(1) {\cal P}
\right\}_{11} - 1\right] ~. \nonumber
\end{eqnarray}
We note that for standard DWF without projection the subtraction and
multiplicative normalization is done automatically, {\it i.e.},
with $\bar q = (\bar \Psi {\cal J} {\cal P})_1$ where ${\cal J}_{ij} =
\delta_{i,L_s+1-j}$ is the inversion operator of the fifth direction,
\be
\< q \bar q \> &\!\!\!\!\!=&\!\!\!\!\! \left\{ {\cal P}^{-1}
[ D^{(5)}_{DW}(m) ]^{-1} {\cal J} {\cal P} \right\}_{11} \nonumber \\
&\!\!\!\!\!=&\!\!\!\!\! \frac{1}{1-m} \left[D_{tov}^{-1}(m) - 1 \right] ~.
\ee

\begin{figure}
\epsfxsize=2.5in
\centerline{\epsfbox[50 60 600 580]{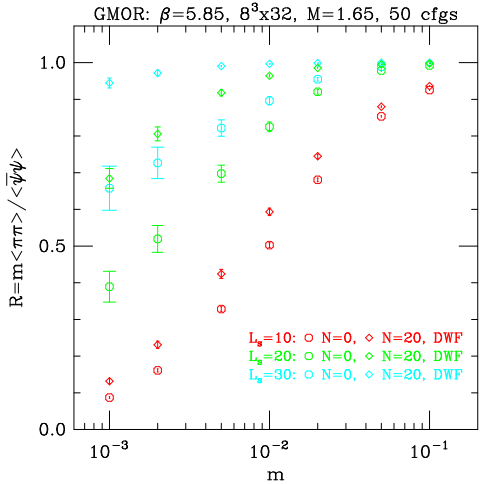}}
\vskip -4mm
\caption{The ratio $R= m \chi_\pi / \< \bar \psi \psi \>$ for DWF without
and with projection of the lowest 20 eigenvectors.}
\vskip -4mm
\label{fig:GMOR}
\end{figure}

Finite $L_s$ and no projection lead to strong violations of the
Gellmann-Oakes-Renner relation (\ref{eq:GMOR}). With projection of a
sufficient number of low eigenvectors the GMOR relation holds down to
much lower quark masses, as seen in Fig.~\ref{fig:GMOR}.

\section{Comparing Efficiencies}

\begin{figure}
\epsfxsize=2.5in
\centerline{\epsfbox[0 50 550 570]{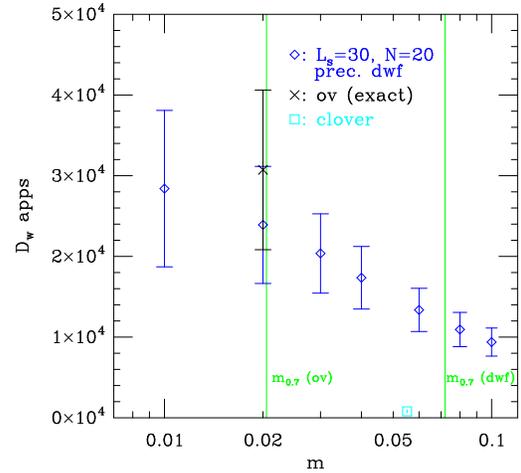}}
\vskip -4mm
\caption{The cost of various implementations of chiral fermions as
measured by the number of applications of $D_w$ on a 4-d vector.}
\vskip -4mm
\label{fig:cost}
\end{figure}

Having different possibilities for chiral fermion formulations on the lattice
we can compare their efficiencies in numerical simulations. For this we
consider a spectroscopy calculation for a topologically trivial gauge
configuration. The cost is measured in the number of times $D_w$, common
to all approaches, has to be applied on a 4-d vector. This is shown in
Fig.~\ref{fig:cost}.

For a single quark mass preconditioned DWF, with projection to assure minimal
induced chiral symmetry breaking (see Fig.~\ref{fig:GMOR}) at negligible
cost, appears most efficient. However, in a quenched spectroscopy calculation
with overlap fermions one can use a multi-shift inverter to compute for all
quark masses simultaneously, compensating for the factor 3-4 in increased
cost. Which regularization to prefer depends thus on the details of the
numerical simulation attempted. If DWF are chosen then the projection method
described here and in \cite{EH_5d} should definitely be used to ensure
good chiral properties at moderate $L_s$.

This research was supported in part by DOE contracts DE-AC05-84ER40150,
DE-FG05-96ER40979 and  DE-FG02-97ER41022.

\end{document}